# The dynamics of sperm cooperation in a competitive environment

Heidi S. Fisher[1,2], Luca Giomi[3,5], Hopi E. Hoekstra[1,2], L. Mahadevan[1,3,4]

[1]Department of Organismic & Evolutionary Biology, [2]Howard Hughes Medical Institute, Department of Molecular & Cellular Biology, Museum of Comparative Zoology, [3]School of Engineering & Applied Sciences, [4]Department of Physics
Harvard University, Cambridge, MA 02138, USA
[5]SISSA, International School for Advanced Studies, Trieste, Italy

## Abstract

Sperm cooperation has evolved in a variety of taxa and is often considered a response to sperm competition, yet the benefit of this form of collective movement remains unclear. Here we use fine-scale imaging and a minimal mathematical model to study sperm aggregation in the rodent genus *Peromyscus*. We demonstrate that as the number of sperm cells in an aggregate increase, the group moves with more persistent linearity but without increasing speed; this benefit, however, is offset in larger aggregates as the geometry of the group forces sperm to swim against one another. The result is a non-monotonic relationship between aggregate size and average velocity with both a theoretically predicted and empirically observed optimum of 6-7 sperm/aggregate. To understand the role of sexual selection in driving these sperm group dynamics, we compared two sister-species with divergent mating systems and find that sperm of *P. maniculatus* (highly promiscuous), which have evolved under intense competition, form optimal-sized aggregates more often than sperm of *P. polionotus* (strictly monogamous), which lack competition. Our combined mathematical and experimental study of coordinated sperm movement reveals the importance of geometry, motion and group size on sperm velocity and suggests how these physical variables interact with evolutionary selective pressures to regulate cooperation in competitive environments.

## 1. Introduction

The factors that contribute to reproductive success are numerous and complex, yet across vertebrates, relative sperm motility is often the best predictor of male fertility [1-7]. When competition among males intensifies, adaptations that improve sperm swimming performance are therefore expected to be strongly favored [8,9]. Indeed, comparisons between related taxa reveal that sperm of polyandrous species, in which females mate with multiple partners during a reproductive cycle, swim faster than sperm from closely-related monogamous species [10,11]. Among the many strategies that improve sperm swimming performance, perhaps the most intriguing mechanism involves cooperation or association with other motile cells [12]. Even without direct attachment, sperm of some species interact with one another via flow fields that result from hydrodynamic interactions [13]. These associations, however, are magnified when multicellular groups form by conjugation, ranging in size from sperm pairs to large aggregates containing hundreds of sperm (reviewed in [14,15]). Sperm aggregation is often assumed to improve motility, yet comparative studies have shown inconsistent results (reviewed in [14,15]), and the underlying mechanics of the associations remain largely unknown.

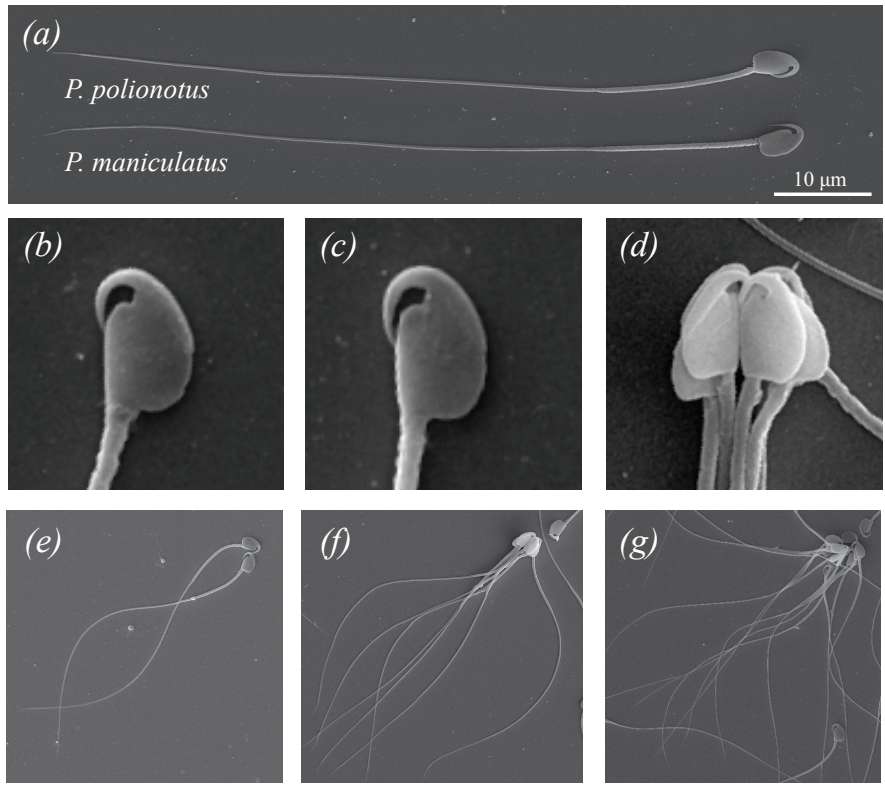

**Figure 1** Scanning electron micrographs of (*a*) whole *Peromyscus* sperm cells, and head morphology of a single (*b*) *P. maniculatus* and (*c*) *P. polionotus* sperm. (*d*) Head orientation of sperm in a typical aggregate with hooks facing inward, and aggregates consisting of (*e*) two, (*f*) seven, and (*g*) thirteen *P. maniculatus* cells.

Like most muroid rodents, sperm from mice in the genus *Peromyscus* typically possess an apical hook on the head (Figure 1a-c)[16] that is thought to facilitate the formation [17] and/or stabilization [18] of sperm groups (but see [19]). Aggregations of *Peromyscus* sperm cells are formed by secondary conjugation [12]: sperm are ejaculated as solitary cells but quickly begin to form multicellular aggregates by adhering to one another at or near the hook (Figure 1d)[20]. Overall these motile sperm groups have a larger average velocity (straight-line velocity [VSL], Figure 2) than single cells; however, the largest groups, those over twenty cells, are often not motile at all [20]. Understanding how sperm aggregates achieve greater average velocity than single cells, whether by increasing their speed (curvilinear velocity [VCL], Figure 2) or traveling in a straighter trajectory (linearity), and how group size can hinder motility, is key to understanding how post-copulatory male-male competition may be acting on sperm behavior to drive and constrain group formation.

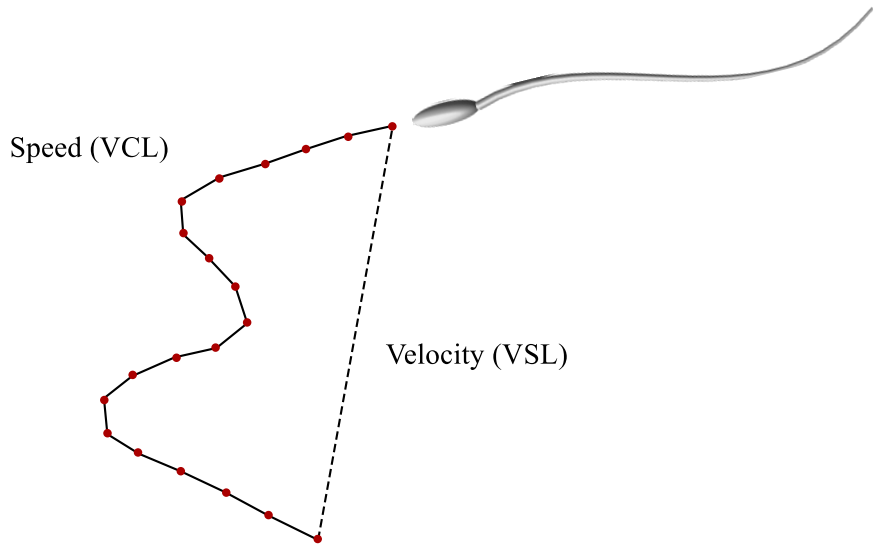

**Figure 2** Schematic representation of the average velocity (VSL) and speed (VCL). VSL is calculated by dividing the distance between the initial and final position in a sperm trajectory (dashed line) by the time **Δ*t*** employed to move; VCL is found by dividing the length of the actual curvilinear trajectory (solid line) by **Δ*t***.

In the genus *Peromyscus*, sperm competition is predicted to be greatest in *P. maniculatus* since both sexes mate with multiple partners, often in overlapping series just minutes apart [21], and females frequently carry multiple-paternity litters in the wild [22]. By contrast, its sister species, *P. polionotus,* is strictly monogamous on the basis of both behavioural [23] and genetic data [24]. The sperm of both species form aggregations with similar geometry and cell orientation, likely due to analogous morphology of their sperm heads [25], yet the competitive environments experienced by *P. maniculatus* and *P. polionotus* sperm represent divergent selective regimes, which is believed to shape how cooperative sperm groups assemble [20]. Here we use a minimal mathematical model to predict how sperm can improve their average velocity by forming aggregations and then use fine-scale imaging to test these predictions and gain a deeper understanding of how sexual selection has acted on this unique form of cooperation in *Peromyscus* sperm.

## 2. Material and methods

### (a) Mathematical model

A simple mechanistic picture of how the average velocity of sperm is a non-monotonic function of aggregate size is suggested by the geometry of the aggregates shown in Figure 1e-g. As sperm cells form small oriented clusters, their motive force and cluster geometry can increase due to the collective beating of their flagella that leads to a greater dynamical persistence. However, in large clusters, the geometry of the aggregate approaches that of an isotropic cluster such that their collective ability to move is severely hindered. A minimal model described below allows us to quantify the advantage of cooperation in a competitive environment using observable physical variables.

Our approach follows a set of models originally developed for flocking behavior of organisms [26,27], which have been used successfully to describe collective motion in a variety of natural and artificial systems, including fish and birds [28], insects [29], bacterial colonies [30] and robots [31] (for details, see Appendix). In this spirit, we treat sperm as individual self-propelled particles [32] that can interact with each other geometrically and mechanically, consistent with the biology of *Peromyscus* sperm aggregation [20]. We restrict our attention to the dynamics of the aggregates once they form, not attempting to address the process of hydrodynamic self – organization itself. Our approach relies on three basic assumptions: (1) although the flagellum is responsible for propulsion, it does not contribute to mechanical interactions between sperm; (2) the main physical mechanism associated with aggregate formation is due to adhesion between sperm heads, consistent with our understanding of sperm morphology [12]; (3) hydrodynamic interactions between sperm in the aggregate are negligible. Thus, although hydrodynamic interactions among neighboring sperm are important in creating self-organized patterns of swimming [33-35], in our minimal model that focuses on the dynamics of the aggregate, these interactions do not play a critical role.

With the aim of characterizing the empirical system using a small number of experimentally measurable parameters, we consider exclusively those features of sperm mechanics that are essential for the formation of motile aggregates. Thus, we note that individual sperm occupy space, are able to move and can link to other sperm. Single sperm cells are then represented as two-dimensional tailless elliptical particles that self-propel at constant velocity $v_0$ in a plane in the direction of their major axis $\mathbf{n}$ while being subject to random planar rotations. Each particle is assumed to have a given number of "keys" and "locks," representing the adhesion complexes on the sperm head. When the key of a particle is within a certain distance $r_a$ from the lock of another particles, a link, represented by a linear spring of stiffness $k_a$, is formed (Figure 3a). If the key-lock distance eventually becomes larger than $r_a$, the link breaks and the two sperm unbind (i.e. an individual adhesion complex can withstand forces up to a stall force $F_a = k_a r_a$). Finally, the particles are themselves assumed to be hard and unable to overlap so that when in contact, they pack as dictated by their geometry.

The behavior described above leads to equations of motion for the position of the $i^{th}$ sperm given by $\mathbf{r}_i(t)$ and its orientation $\theta_i(t)$ relative to the $x$−axis of the lab frame given by:

$$\frac{d\boldsymbol{r}_i}{dt} = v_0 \boldsymbol{n}_i + \mu^{-1}\boldsymbol{F}_i, \qquad \frac{d\theta_i}{dt} = \zeta_i + \gamma^{-1}M_i \qquad (1)$$

where the $i^{th}$ sperm has its major axis along $\boldsymbol{n}_i = (\cos\theta_i, \sin\theta_i)$, $\boldsymbol{F}_i$ is the total force acting on the $i^{th}$ particle resulting from the short-range steric interactions with the neighbors and adhesion:

$$\boldsymbol{F}_i = k_s \sum_{j=1}^{N_i} \delta_{ij}\boldsymbol{N}_{ij} + k_a \sum_{j=1}^{L_i} \ell_{ij}\boldsymbol{L}_{ij} \qquad (2)$$

and $M_i$ is total torque acting on the $i^{th}$ particle:

$$M_i = \sum_{j=1}^{N_i+L_i} (\boldsymbol{d}_{ij}\times\boldsymbol{f}_{ij}) \cdot \hat{\boldsymbol{z}} \qquad (4)$$

where $N_i$ is the number of neighbors of the $i$−th particle, $L_i$ is the number of adhesive links, $k_s$ is the elastic constant associated with steric interactions, $k_a$ is the adhesive spring elastic constant, with $k_s >> k_a$. Furthermore, $\delta_{ij}$ and $\ell_{ij}$ representing the length of the springs associated with the steric and adhesive interactions, with $N_{ij}$ and $\boldsymbol{L}_{ij}$ unit vectors in the direction of the springs, $\mu$ and $\gamma$ are translational and rotational drag coefficients, $\boldsymbol{f}_{ij}$ is any of the force exerted between the $i^{th}$ and $j^{th}$ cell appearing in Eq. (2) and $\boldsymbol{d}_{ij}$ the associated lever arm, $\hat{\boldsymbol{z}}$ is the normal to the 2-dimensional plane of motion, and $\zeta_i$ is a zero mean delta-correlated Gaussian random variable:

$$\langle \zeta(t)\zeta(t')\rangle = 2D\delta(t-t') \qquad (3)$$

where $D$ is a rotational diffusion coefficient. Here we have assumed that the motion of the sperm is inertialess, consistent with the low Reynolds number regime they operate in, and further have ignored the effect of randomness in the translational degrees of freedom for simplicity.

Our minimal mechanistic model of interacting sperm captures the geometry of the individual sperm, their autonomous movement, and finally their ability to interact with each other adhesively without overlap. While there are many possible variants of these models, the critical parameters in all of them will be qualitatively similar: the aspect ratio of the sperm head, the scaled ratio of the rotational Brownian motion to the interaction torque between cells, the scaled ratio of the adhesive bond strength to random fluctuations, and the relative orientation of the adhesive bonds. These parameters together characterize the dynamics and persistence of movement in aggregates.

## (b) Sperm imaging and analysis

Captive stocks of wild-derived *Peromyscus maniculatus bairdii* and *Peromyscus polionotus subgriseus* were obtained originally from the Peromyscus Genetic Stock Center and have been maintained at the Harvard University in accordance with guidelines established by Harvard's Institutional Animal Care and Use Committee. We used adult (age>90 days) sexually mature *P. polionotus* (*n*=9) and *P. maniculatus* (*n*=9) males for cross-species comparisons.

After sacrifice, we immediately removed the left caudal epididymis of each male, made a single small incision in the tissue, submersed it in 1mL of warmed Biggers-Whitten-Whittingham media [36], and incubated the tissue for 10 min at 37°C to release motile sperm. After the 10 min incubation, we removed the epidydimal tissue, gently swirled the media and incubated for another 5 min. We collected 20μl of media containing live sperm just below the surface of the aliquot, to reduce the number of dead cells, which sink to the bottom. We placed the aliquot on a plastic microscope slide and covered the sample with a plastic coverslip (plastic reduces adhesion of sperm to the slide compared with glass products), and recorded three 5 sec videos (30 frames/sec) of live sperm at 100X magnification under phase contrast conditions on an upright microscope (AxioImager.A1, Zeiss, Jena, Germany).

To examine the dynamic performance of sperm aggregates, we quantified the speed and velocity of both single cells and aggregated groups. The speed, also referred to as curvilinear velocity (VCL), characterizes the rate of change of the two-dimensional projection of an aggregate's trajectory over time (Figure 2). The average velocity, also referred to as straight-line velocity (VSL), is defined as the rate of change of the projected distance along the vector connecting the initial and final point in the trajectory (Figure 2). We acquired VSL and VCL data from video using the Computer Assisted Sperm Analyzer plugin for NIH ImageJ [37], which tracks motile sperm cells or groups to calculate VSL and VCL. We then estimated average linearity (VSL/VCL) for each track. Specifically, for each video recorder we first used the "Find edges" and "Threshold" functions to isolate sperm images from the background and imposed a filter to discard tracks with VSL<5μm/sec or VCL<25μm/sec (cutoffs imposed to avoid non-progressively motile sperm cells or groups). We then used the first 50 tracks (including both single sperm cells and sperm groups) recorded from each donor male in subsequent analyses for all but two males: In the case of one male of each species, fewer than 50 tracks met our criteria (*P. maniculatus* male, *n*=30 tracks; *P. polionotus* male, *n*=27 tracks). Sperm group size was then subsequently counted for each track and verified on at least 5 different frames per track.

We used two-factor (group size, donor male) two-tail ANOVAs to assess the effect of each factor on sperm average velocity (VSL), speed (VCL) and linearity (VSL/VCL) within each species. After identifying the sperm aggregate size that achieved the greatest average velocity (*n*=7 cells), we then compared the average velocity of seven-celled aggregates (the null) to the average velocity of all other sizes for each species using a one-sample two-tailed t-test. Next, we split the *P. maniculatus* and *P. polionotus* data into two groups and used a linear regression (with donor male as a covariate) to test the significant relationship between: group size and average velocity at or below the optimum (*n*≤7 cells), and above the optimum (*n*>8 cells). To identify how sperm aggregate size varies between species, we first averaged group size over each donor male, then used a two-sample two-tailed t-test to compare means, and an F-test to compare variances, of *P. maniculatus* and *P. polionotus* sperm aggregates. Finally, we used a 2-way ANOVA (species, donor male) to compare difference between average linearity achieved by *P. maniculatus* and *P. polionotus* males. All statistical analyses were performed in R [38].

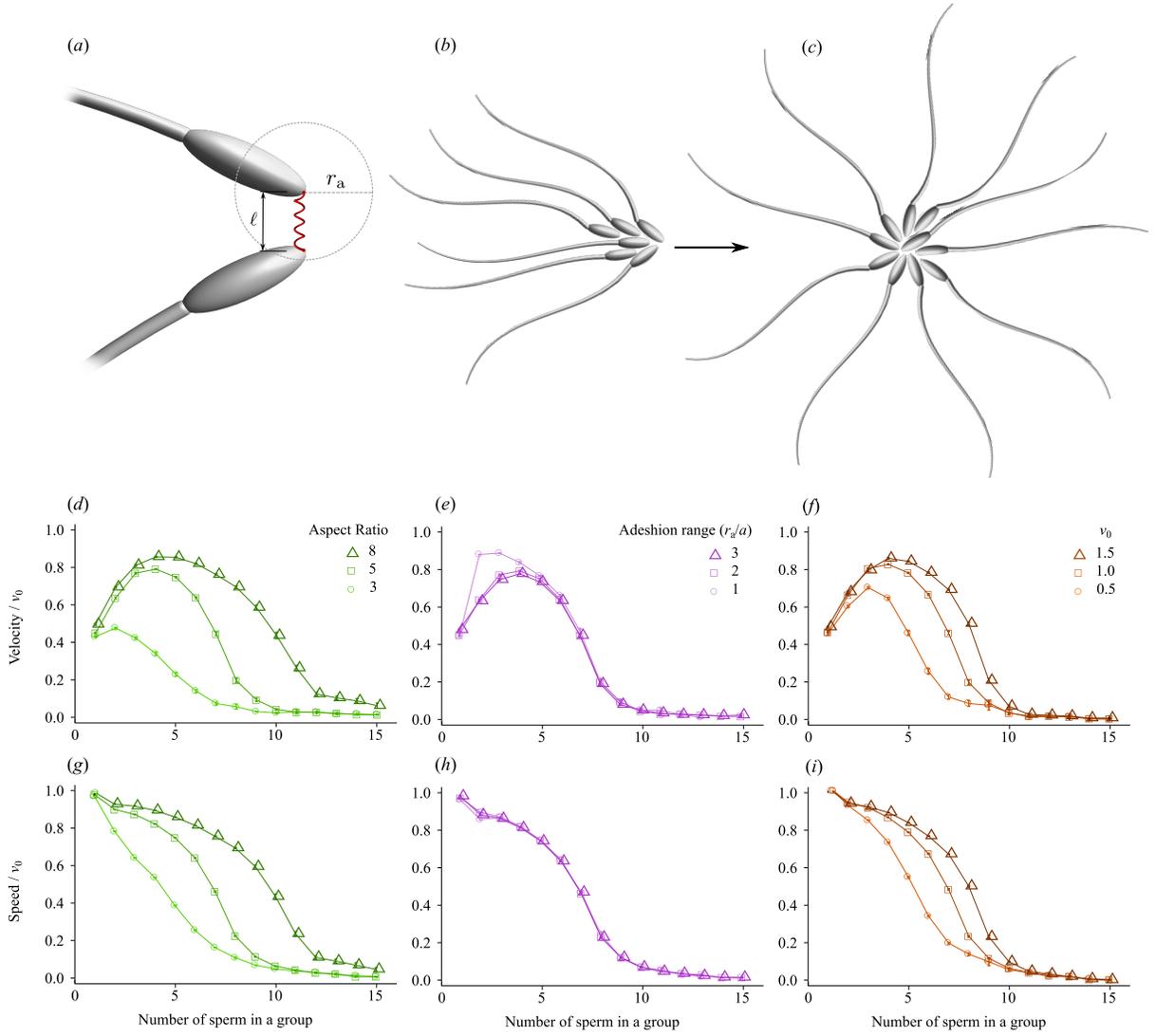

**Figure 3** (*a*) Schematic representation of the adhesive interactions modeled in Eq. (2). Sperm heads are treated as self-propelled elliptical particles whose major and minor semi-axes have length *a* and *b* respectively. Each particle is equipped with a given numbers of keys and locks, representing the adhesion complexes where the sperm can bind. When the key of a particle is within a certain distance $r_a$ from the lock of another particle, a link represented by a linear spring is formed. The geometry of the aggregates affects their motility, so asymmetric aggregates (*b*) move fast and maintain a straight trajectory, while star-shaped aggregates (*c*) move slowly because the velocity of the individual cells in the aggregate cancels each other. Average velocity (*d-f*) and speed (*g-i*) versus aggregate size obtained from a numerical integration of Eq. (1) for various aspect ratios *a/b* (*d*, *g*), adhesion range ranges $r_a$ (*e*, *h*), expressed in units of the particles major semi-axis length *a*, and propulsion velocity $v_0$, in units of $a/\tau_c$ (with $\tau_c = 1/D$ is the time scale associated with the rotational noise) (*f*, *i*).

# 3. Results

## (a) Mathematical model

We integrated Eqs. (1) numerically for a wide range of parameter values. Our model sample consists of 100 cells in a square domain of size $L$=500 (in units of the particle minor semi-axis $b$) with periodic boundary. For all choices of parameters, aggregation always leads to a prominent increase in the average velocity (but not speed) for small aggregate size, while large aggregates suffer from both reduced velocity and speed (Figure 3d-i). The origin of this behavior can be explained by noting that sperm can associate with each other via soft adhesive bonds, modeled here as finitely extensible springs (see Section 2a). Once they are linked, they form aggregates whose structure is predominantly dictated by the geometry and the spatial distribution of the adhesive patches. The structure of the aggregates affects how the velocity of the individual sperm determines the final velocity of the aggregates. Thus, radially symmetric aggregates consisting of many sperm (e.g. Figure 3c) are likely to be non-motile because the velocities of the individual cells effectively cancel each other. Smaller aggregates, on the other hand, are asymmetric and maintain the typical head/tail directionality of individual sperm (e.g. Figure 3b). More importantly, their close packed structure reduces the random fluctuations in the swimming direction of the individual cells, resulting in a persistent linearity of the trajectory and therefore a higher average velocity (VSL). The combined effect of these two competing mechanisms leads to an optimal aggregate size.

The precise value of the optimal aggregate size, as well as the sharpness of the velocity peak, depends on the detailed geometry of the head/mid-piece complex and the adhesion properties of the sperm heads. To investigate how cell geometry affects the swimming performance of an aggregate, we simulated self-propelled particles of various aspect ratios, the ratio between the length of the major and minor semi-axes of the elliptical particle. Increasing the slenderness of the particles moves the velocity optimum toward larger aggregate sizes and simultaneously reduces the slope of the speed curve (Figure 3d, g). This is because slender elliptical particles can pack more tightly than circles in two dimensions, so that it requires a larger number of particles to reach a symmetric conformation. Increasing the adhesion range $r_a$ (thus the stall force that a single adhesive bond can withstand) also has the effect of moving the optimum toward larger aggregates (Figure 3e), while leaving the speed essentially unaltered (Figure 3h). Increasing the sperm propulsion velocity $v_0$ (Figure 3f) affects the position of the optimum only slightly, but produces a substantial improvement in the dynamic performance of aggregate.

Finally, we note that in our two-dimensional self-propelled particles model, the aggregate size at which the speed starts to drop has a straightforward geometrical interpretation related to the *kissing number* of the particles, defined as the number of particles that can touch a given central particle without overlap. If the ellipses are not excessively slender, this number equals six (the same as for circles), thus aggregates formed by six or more ellipses tend to be highly symmetric and undergo a severe drop in speed (Figure 3g-i). *Peromyscus* sperm cells have a flat head-shape roughly similar to a very oblate ellipsoid (Figure 1b,c). For this type of shapes one might expect a kissing number between six and twelve, the latter being the kissing number for spheres in three dimensions.

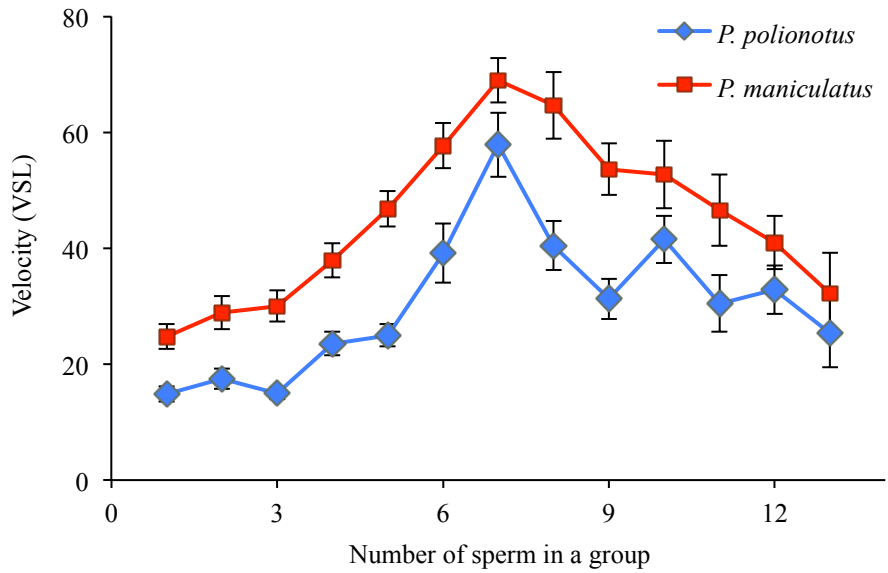

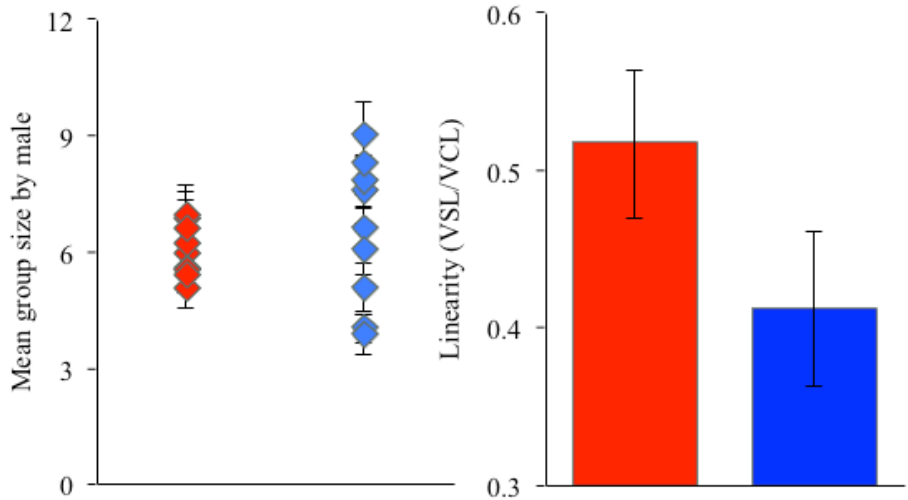

**Figure 4** Size and performance of *Peromyscus* sperm aggregates. (*a*) Mean ± standard error (SE) of velocity (VSL) of sperm aggregates by group size over all donor males from each species. (*b*) Mean ± SE group size of aggregated sperm in each species, sperm from each donor is represented as a separate point with error bars. (*c*) Mean ± SE linearity (VSL/VCL) of aggregated sperm over all males from each species; note truncated y-axis.

## (b) Experiments

In both *P. maniculatus* and *P. polionotus,* motile sperm groups varied in size, ranging from 1-35 cells/group. We found that, overall, group size significantly influences average velocity (VSL) in both species, even after the variation between donor males is accounted for (Table 1). However, there is no significant relationship between the number of sperm in a group and speed (VCL), yet similar to the result for average velocity, we found a significant effect of donor male in both species for speed (Table 1). Finally, when we measured the linearity (VSL/VCL) of all sperm groups, we found that group size significantly effects linearity, with donor male as a covariate, in both species (Table 1).

| Measure | Species | Factor | F | df | p-value |
|---|---|---|---|---|---|
| **Average velocity (VSL)** | *P. maniculatus* | group size | 34.3 | 1 | $9.41 \times 10^{-9}$ |
| | | donor male | 16.7 | 8 | $<1 \times 10^{-15}$ |
| | *P. polionotus* | group size | 42.1 | 1 | $2.52 \times 10^{-10}$ |
| | | donor male | 7.6 | 8 | $1.90 \times 10^{-9}$ |
| **Speed (VCL)** | *P. maniculatus* | group size | 2.1 | 1 | 0.15 |
| | | donor male | 5.7 | 8 | $6.82 \times 10^{-7}$ |
| | *P. polionotus* | group size | 3.5 | 1 | 0.06 |
| | | donor male | 4.9 | 8 | $8.41 \times 10^{-6}$ |
| **Linearity (VSL/VCL)** | *P. maniculatus* | group size | 89.4 | 1 | $<1 \times 10^{-15}$ |
| | | donor male | 17.4 | 8 | $<1 \times 10^{-15}$ |
| | *P. polionotus* | group size | 131.4 | 1 | $<1 \times 10^{-15}$ |
| | | donor male | 16.5 | 8 | $<1 \times 10^{-15}$ |

**Table 1** Two-factor two-tailed ANOVAs on sperm performance data.

The greatest average velocity was achieved by groups of seven sperm cells (Figure 4a) and aggregates both smaller ($n<6$ cells) and larger ($n>8$ cells) than this number were slower in both species (*P. maniculatus* $t=4.2$, $df=8$, $p=0.003$; *P. polionotus* $t=10.4$, $df=8$, $p=0.0001$). Moreover, we found a significant positive association between sperm aggregate size and average velocity in both species as group size increased from 1 to 7 sperm cells (*P. maniculatus* $R^2=0.48$, $p=2.0 \times 10^{-16}$; *P. polionotus* $R^2=0.42$, $p=2.0 \times 10^{-16}$), yet a significant decrease as groups grew larger than 7 cells (*P. maniculatus* $R^2=0.39$, $p=1.63 \times 10^{-10}$; *P. polionotus* $R^2=0.14$, $p=5.20 \times 10^{-4}$).

When we averaged sperm performance for each male in both species, we found that mean aggregate size did not differ significantly between species (mean±SE=cells/group: *P. maniculatus*=6.0±0.2 range=2-26 cells, *P. polionotus*=6.5±0.72 range=2-31 cells, $n=50$ groups/male, $n=9$ males; $p=0.51$, $df=8$); however, the average group size in *P. polionotus* is significantly more variable than in *P. maniculatus* (Figure 4b; F-test: $p=0.044$). Moreover, the average linearity (VSL/VCL) achieved by sperm of *P. maniculatus* males is significantly greater than the average linearity of *P. polionotus* sperm (Figure 4c; $F=47.45$, $df=1$, $p=1.11 \times 10^{-11}$).

## 4. Discussion

Our combined theoretical and experimental approach, allowed us to build a mathematical model based on biological observations with testable predictions. Our empirical results are consistent with the first two salient predictions of our model: (1) when sperm conjugate in a head to head formation, such as in *Peromyscus*, and when aggregate size exceeds the optimum, cells will exert opposing forces upon one another and thereby reduce the velocity of the entire group, and (2) the optimal size is dictated largely by the geometry of the sperm heads, and therefore species with similar sperm heads, such as *P. maniculatus* and *P. polionotus* [25], will achieve the same optima. We also found that, overall, group size significantly influences average velocity (VSL) in both species, and the greatest average velocity is achieved by groups of seven sperm cells— aggregates both smaller ($n<6$ cells) and larger ($n>8$ cells) than this number are progressively slower in both species. Taken together our results suggest that the shared aggregate geometry of *P. maniculatus* and *P. polionotus* sperm (likely as a result of the similarly shaped [Figure 1b,c] and sized [25] sperm heads) results in a similar relationship between sperm group size and performance, and thus similar optima, in these species.

A third prediction of the model is that sperm aggregates achieve greater average velocity (VSL) because they move in a more linear path, rather than an increase in speed (VCL). Indeed, we found no significant relationship between the number of sperm in a group and speed in either *P. maniculatus* or *P. polionotus*. However, like velocity, we found a significant effect of donor male in both species on speed; this variation among males and between species is consistent with earlier findings of inter-male differences in speed in these species [39]. In contrast, when we measured the linearity (VSL/VCL) of all sperm groups, we found a significant effect of group size on linearity with donor male as a covariate in both species. These results indicate that the benefit of sperm aggregation is, indeed, conferred via a more direct path of travel, rather than a change in speed, as predicted by our model.

Our experimental results are consistent with the predictions of the model that shared aggregate geometry of *P. maniculatus* and *P. polionotus* will yield similar relationships between sperm group size and performance in both species. In nature, however, the ideal strategy is not always the most prevalent one due to associated costs, selection on pleiotropic traits, and/or genetic drift. While *P. maniculatus* sperm have evolved under a selective regime with intense competition [21,22], evidence suggests that monogamy in *P. polionotus* [23,24] is derived [40] and, therefore, sexual selection is likely relaxed in *P. polionotus*. When we measured the average aggregate size in each male across the two species, we found that while the average group size does not differ significantly between species, the average group size in *P. polionotus* is significantly more variable than *P. maniculatus*. In other words, the mean group size does not differ between the species, which are both within one cell of the observed (empirical) and predicted (theoretical) optimum, but the distribution around the mean is significantly larger in *P. polionotus,* and thus more aggregates are further away from the optimum, compared with *P. maniculatus*. These results suggest that sexual selection, and male-male competition specifically, may be imposing stabilizing selection on aggregate size in *P. maniculatus* sperm, resulting in more groups at or near optimal size; by contrast, the monogamous mating system of *P. polionotus* represents a relaxation of male competition and is consistent with greater variation in sperm group size.

Given that *P. maniculatus* sperm are more likely to form aggregates at or near the optimal size compared to *P. polionotus*, our model also predicts an overall increased linearity in *P. maniculatus* sperm in the total sample. Indeed, *P. maniculatus* sperm move in a more direct trajectory (VSL/VCL) than *P. polionotus* sperm. The results from this study reveal that selection may, therefore, act on sperm swimming performance via aggregation behavior.

## 5. Conclusions

Our detailed observations of sperm shape, aggregate geometry, and their dynamical performance suggest an optimal sperm aggregate size that leads to a maximum linear velocity of a group. Our minimal mathematical model—that accounts for the geometry of the sperm, the mechanics of their adhesive interactions, and combined with the dynamics of their fluctuating translational and rotational movement—captures the non-monotonic dependence of aggregate velocity on the number of sperm in a group. The underlying mechanism is simple: in small groups, sperm adhesion reduces the size of rotational fluctuations by effectively canceling them, while in large aggregates, this effect is dominated eventually by reducing the mean translational velocity due to the isotropic geometry of a cluster. Thus, relatively few mechanical parameters can explain the dynamics of a seemingly complex biological process.

Our empirical data test these model predictions and show that sperm achieve greater velocity, surprisingly not by increasing speed, but rather by traveling in a more direct path than solitary cells. This collective behavior arises from direct physical interaction among cells, which determines the optimal aggregate size. The number of cells involved in an aggregate, therefore, greatly contributes to sperm performance and the reproductive success of a male, thereby offering another dimension of sperm biology on which selection can act. Moreover, by comparing sperm dynamics in two species that have evolved under disparate competitive regimes, we are able to implicate a role for sexual selection in the evolution of complex behavior of these seemingly simple cells.

Thus a deep understanding of sperm behavior requires us to combine our knowledge of geometrical and physical constraints with reproductive biology; indeed these dynamics are clearly driven by a combination of morphology, kinematics and the selective environment. While selection ultimately acts on organismal fitness, our picture allows us to link this to the dynamics of movement and the adhesive interactions among sperm. Indeed, sperm cooperation and competition are not only a remarkable arena to study evolutionary processes but also to test quantitative models for how they may play out in nature.

**Acknowledgements.** We thank James Weaver for generating the scanning electron microscopy images. This research was funded by an N.I.H. Ruth Kirschstein National Research Service Award and an N.I.H. Pathway to Independence Award to H.S.F., an Arnold and Mabel Beckman Foundation Young Investigator Award and funds from the Howard Hughes Medical Institute to H.E.H., and a MacArthur Fellowship to L.M.

# Appendix

### (a) Single sperm dynamics

To understand why aggregation improves the mobility of sperm, it is useful to consider the dynamics of a single cell resulting from Eq. (1) in the main text:

$$\frac{d\mathbf{r}}{dt} = v_0 \mathbf{n}, \qquad \frac{d\theta}{dt} = \zeta, \tag{4}$$

Where $\zeta$ is a zero mean delta-correlated Gaussian random variable: $\langle \zeta(t)\zeta(t')\rangle = 2D\delta(t-t')$, with $D$ a rotational diffusion constant. Eqs. (4) describe the motion of a cell moving at constant speed $v_0$, but whose direction of motion is affected by random rotations. Translational noise can be neglected due to the fact that the typical Peclect number of a swimming sperm of size $L$ is $Pe = v_0 L/D = 10^4 - 10^5$ [41]; thus diffusion is negligible compared to drift. The beating of the tail that controls the orientation of the cell, on the other hand, is subject to fluctuations due to the noise in the activity of the motors regulating the flagellar beating. Eqs. (4) can be easily solved using the standard machinery of Brownian motion (see for instance [42,43]). Integrating the $\theta$ equation yields:

$$\Delta\theta(t) = \theta(t) - \theta(0) = \int_0^t dt' \zeta(t'), \tag{5}$$

from which:

$$\langle \Delta\theta(t)\rangle = 0, \qquad \langle \Delta\theta(t)\Delta\theta(t')\rangle = 2D \ \min(t,t') \tag{6}$$

where min(*t*,*t'*) represent the minimum between the times *t* and *t'*. Now, as $\Delta\theta$ is a linear combination of random variable with zero mean and variance $\langle \Delta\theta^2(t)\rangle = 2Dt$, from the central limit theorem it follows that it is Gaussianly distributed. Namely:

$$P[\Delta\theta(t)] = \frac{1}{\sqrt{4\pi Dt}} e^{-\frac{\Delta\theta^2(t)}{4Dt}}, \tag{7}$$

where we called $P[\Delta\theta(t)]$ the probability density function associated with $\Delta\theta(t)$. This allows us to calculate the averages of exponential and trigonometric functions:

$$\langle e^{\pm i\Delta\theta(t)}\rangle = e^{-\frac{1}{2}\langle \Delta\theta^2(t)\rangle} = e^{-Dt}, \tag{8}$$

from which:

$$\langle \cos\theta(t) \rangle = e^{-Dt}\cos\theta_0, \qquad \langle \sin\theta(t) \rangle = e^{-Dt}\sin\theta_0. \tag{9}$$

Averaging both sides of the position equation (4) allows us to calculate the average velocity of a sperm cell:

$$\langle \mathbf{v}(t) \rangle = v_0 e^{-Dt}\mathbf{n}(0), \tag{10}$$

where $\mathbf{n}(0) = (\cos\theta_0, \sin\theta_0)$ is the direction of the cell axis at $t = 0$. Eq. (10) represents an important property of our model sperm cell: at short times, individual sperm move ballistically in the direction of their axis, but, due the noisy flagellar beating, their velocity exponentially loses directional correlation and in a time scale of order $\tau_c = 1/D$ the cell has completely lost track of its initial direction. Aggregation allows groups of cells to reduce the fluctuations in the direction of motion, thus increasing the correlation time $\tau_c$ (see later). The mean-square displacement of a single cell can be calculated as:

$$\langle |\mathbf{r}(t) - \mathbf{r}(0)|^2 \rangle = v_0^2 \int_0^t \int_0^t dt_1 dt_2 \, \langle \cos[\theta(t_1) - \theta(t_2)] \rangle, \tag{11}$$

and yields:

$$\langle \cos[\theta(t_1) - \theta(t_2)] \rangle = e^{-\frac{1}{2}\langle [\Delta\theta(t_1) - \Delta\theta(t_2)]^2 \rangle} = e^{-D[t_1 + t_2 - 2\min(t_1, t_2)]} = e^{-D|t_1 - t_2|}. \tag{12}$$

Combining this with Eq. (11) gives:

$$\langle |\mathbf{r}(t) - \mathbf{r}(0)|^2 \rangle = \frac{2v_0^2}{D}\left(t - \frac{1 - e^{-Dt}}{D}\right). \tag{13}$$

Thus for $t \gg \tau_c$ the motion is completely Brownian with $\langle |\mathbf{r}(t) - \mathbf{r}(0)|^2 \rangle \approx 2v_0 t/D$. The average speed (VCL) and velocity (VSL) (Figure 2) of a single cell are then given as follows:

$$\text{VCL} = v_0, \qquad \text{VSL} = \frac{\sqrt{\langle |\mathbf{r}(t) - \mathbf{r}(0)|^2 \rangle}}{t}. \tag{14}$$

Because the mean-square displacement does not depend linearly on time, VSL depends in general on the time range used to calculate it and vanishes in the limit $t \to \infty$, indicating that a cell has been diffusing long enough to visit its initial position more than once. For the dynamics of a single sperm cell, we can replace $t$ in Eq. (12) with the lifetime $\tau_1$ of an isolated cell, defined as the time it takes for the cell to meet other cells and form and aggregate. If the sample consists of a density $\rho$ of uniformly distributed cells, this is approximately given by $\tau_1 \approx 1/(v_o\sqrt{\rho})$. Figure 5 shows the average velocity and the lifetime of isolated cells for various $v_0$ and cell number obtained from numerical integration of Eq. (4).

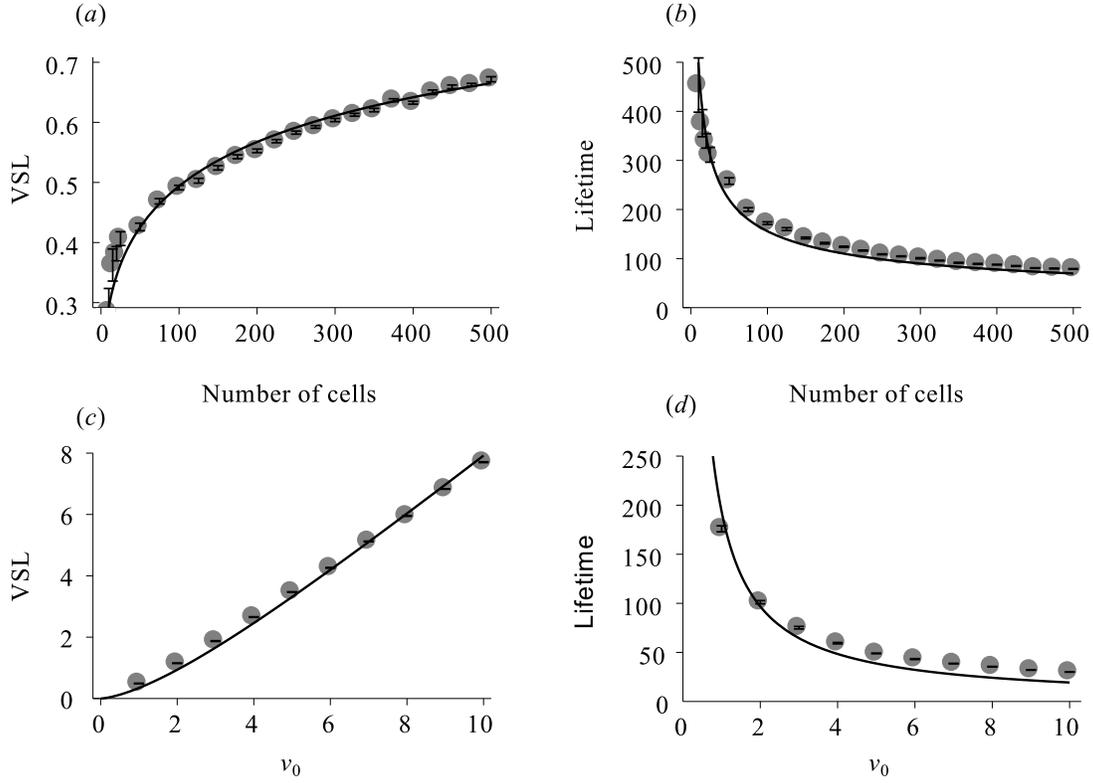

**Figure 5** Average velocity (VSL) and lifetime (i.e. aggregation time) of a single cell. Both VSL and $v_0$ are expressed in units of $a/\tau_c$, with a the length of the cell minor semi-axis and $\tau_c = 1/D$ the correlation time. The circles are obtained by numerical integration of Eq. (4), while the solid lines are given by Eq. (14) by setting $t = A/(v_0\sqrt{\rho})$, with $\rho$ the particle density and $A$ a fitting parameter.

In the experiments, the time range $t_{\exp}$ used to calculate VSL is fixed and is based on the frame rate of the images (30 frames/second). This can be incorporated in the definition of $D$, by introducing a dimensionless rotational diffusion number $\mathcal{D} = D t_{\exp}$. The long time average velocity is then characterized by two quantities: $\text{VCL} = v_0$ and $\text{VSL/VCL} \approx 1/\sqrt{\mathcal{D}}$. Furthermore the discussion presented above is not restricted to the case of an isolated cell, but can be extended to aggregates as well. In the case of an aggregate, VCL is the average velocity of the aggregate, while VSL/VCL can be taken as a measure of the effective diffusion number $\mathcal{D}_{\text{eff}}$ that accounts also for the reduction in the orientational noise resulting from aggregation.

## (b) Scaling

The scaling behavior of the straight-line velocity for increasing aggregate size can be predicted within a mean-field framework starting from a simple geometric argument. As explained in the previous section: $\text{VSL} \sim 1/\sqrt{D} \sim 1/\Delta\theta$. Thus if the angular span $\Delta\theta$ is reduced by a factor $x < 1$, VSL is increased by a factor $1/x$. Now, in the interior of an aggregate $\Delta\theta$ decreases due to the crowded environment. To account for this effect we use the following *ansatz*:

$$\Delta\theta \sim \Omega_d \left(1 - \frac{n}{k_d}\right) \quad (15)$$

where $\Omega_d$ is the solid angle in $d$ dimensions, $k_d$ the $d$-dimensional kissing number and $n$ the average number of neighbors in the aggregate.. Eq. (15) implies that each new neighbor takes an equal amount of angular space until, for $n = k_d$, there is no space left and the cells have reached a jammed configuration.

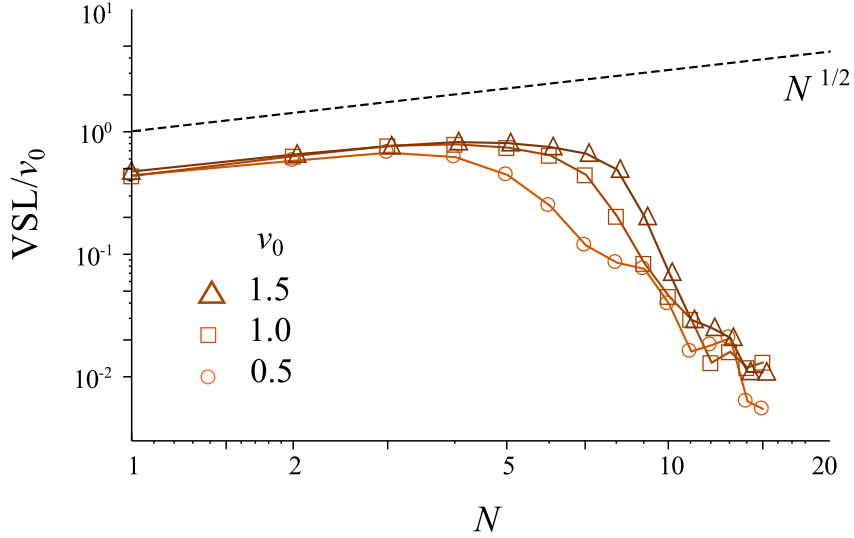

**Figure 6** Log-log plot of the average velocity (VSL) versus aggregate size for various propulsion velocity $v_0$. The numerical data are the same of Figure 3f, while the dashed line shows the predicted scaling law **VSL** $\sim N^{1/2}$ in two dimensions.

The number of neighbors $n$ depends in general on the size $N$ of the aggregate and in particular on the number of cells in the interior of the aggregate compared to those distributed along the boundary. Thus, calling $z_i$ and $z_b$ the average number of neighbors in the interior and on the boundary, we have:

$$n = \frac{z_i N_i + z_b N_b}{N}, \quad (16)$$

where $N_i$ and $N_b$ represent respectively the number of cells in the interior and on the boundary and $N = N_i + N_b$. Next, assuming $z_i = k_d$ and taking into account that $N_b \sim N^{1/d}$ we obtain:

$$n = k_d - c(k_d - z_b)N^{-\frac{d-1}{d}}. \quad (17)$$

with $c$ a constant. Now, when the aggregate consists of a single cell, $n = 0$. This allows to calculate the constant $c$ in the form $c = k_d/(k_d - z_b)$, from which $n = k_d\left(1 - N^{-(d-1)/d}\right)$ and Eq. 15 becomes:

$$\Delta\theta \sim \Omega_d N^{-\frac{d-1}{d}}. \tag{18}$$

Consequently:

$$\text{VSL}(N) = \text{VSL}(1)\, N^{\frac{d-1}{d}} \sim \begin{cases} N^{1/2} & d = 2 \\ N^{2/3} & d = 3 \end{cases} \tag{18}$$

For our simple two-dimensional model this prediction is consistent with the numerical data shown in Figure 6.

## Supplemental figure

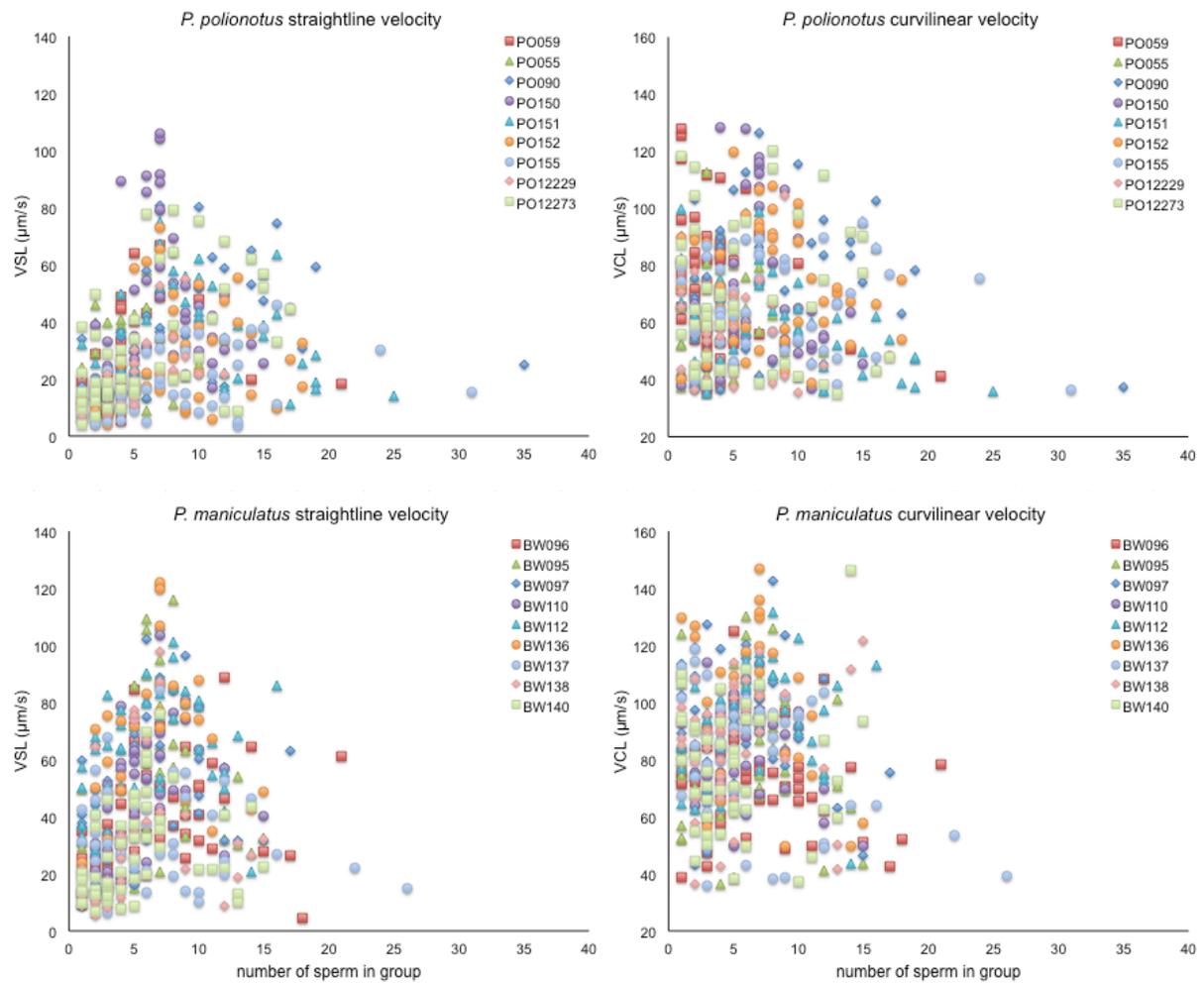

**Figure S1:** Straightline velocity (VSL) and curilinear velocity (VCL) of all sperm aggregates measured, labelled by donor male and indicated by the colored points.